\documentclass[12pt]{article}
\usepackage{mathpazo}
\usepackage[T1]{fontenc}
\usepackage[utf8]{inputenc}

\usepackage{dcolumn}
\usepackage{bm}
\usepackage{braket}
\usepackage{tabularx}
 \usepackage{booktabs}
 \usepackage{enumitem}

\usepackage{xcolor}
\usepackage{amsmath}
\usepackage{bm}
\usepackage{relsize}
\usepackage{comment}
\usepackage{graphicx}
\usepackage[UKenglish]{babel}
\usepackage{microtype}                      
\usepackage{xcolor} 
\usepackage{hyperref}                   	
\usepackage{lipsum}
\usepackage{fancyhdr}
\usepackage{picture} 
\hypersetup{								
	colorlinks=true,      		            
	linkcolor=blue,                        	
	filecolor=blue,                     	
	citecolor=blue,    	        			
	urlcolor=blue,                 	   		
	pdffitwindow=false,               	 	
	pdfstartview={FitH},               		
	pdfauthor={Huw Price},           		
}

\usepackage{braket}
\usepackage{physics}
\usepackage{todonotes}
\usepackage{enumitem}

\usepackage{tikz}
\usetikzlibrary{shapes,decorations,arrows,calc,arrows.meta,fit,positioning}
\tikzset{
    -Latex,auto,node distance =1 cm and 1 cm,semithick,
    state/.style ={ellipse, draw, minimum width = 0.7 cm},
    point/.style = {circle, draw, inner sep=0.04cm,fill,node contents={}},
    bidirected/.style={Latex-Latex,dashed},
    el/.style = {inner sep=2pt, align=left, sloped}
}

\begin{document}

\hypertarget{why-entanglement}{%
\title{A Mechanism for Entanglement?}\label{bell}}
\author{Huw Price\thanks{Trinity College, Cambridge, UK; email \href{mailto:hp331@cam.ac.uk}{hp331@cam.ac.uk}.} {\ and} Ken Wharton\thanks{Department of Physics and Astronomy, San Jos\'{e} State University, San Jos\'{e}, CA 95192-0106, USA; email \href{mailto:kenneth.wharton@sjsu.edu}{kenneth.wharton@sjsu.edu}.}}
\date{\today}
\maketitle\thispagestyle{empty}

\begin{abstract}
\noindent We propose that quantum entanglement is a special sort of selection artefact, explicable as a combination of (i) collider bias and (ii) a boundary constraint on the collider variable. We show that the proposal is valid for a special class of (`W-shaped') Bell experiments involving delayed-choice entanglement swapping, and argue that it can be extended to the ordinary (`V-shaped') case. The proposal requires no direct causal influence outside lightcones, and may hence  offer a way to reconcile Bell nonlocality and relativity. 

\end{abstract}

\section{Introduction}

Quantum entanglement was  first clearly identified, and named,
by Erwin Schr\"odinger in 1935  \cite{Sch35a,Sch35b}. Schr\"odinger was responding to a now-famous paper by Einstein, Podolsky and Rosen (EPR) \cite{EPR}. 
He notes that in the kind of two-part quantum systems that EPR  discuss, two components that have just interacted cannot be described independently, in the way that classical physics would have allowed. As he says: \begin{quote}When two separated bodies that each are maximally known come to interact, and then separate again, then such an \textit{entanglement} of knowledge often happens. \cite[\S10]{Sch35a}\end{quote}
He emphasises the centrality of this point to the new quantum theory:
\begin{quote}
I would not call that \textit{one} but rather \textit{the} characteristic trait of quantum mechanics, the one that enforces its entire departure from classical lines of thought.   \cite[555]{Sch35b}  
\end{quote}

Like EPR, Schr\"odinger thought that entanglement implied that quantum mechanics (QM) as it stood was incomplete. He thought that there must be further properties of the two subsystems, not present in the QM description itself, to explain this `entanglement of knowledge'. Endorsing the EPR argument to the same conclusion, Schr\"odinger says that   the alternative would be that a measurement on one particle affects the remote particle in some way, and he thought that was absurd: ‘Measurements on separated systems cannot affect one another directly, that would be magic.’ EPR explicitly assume that this does not happen, an assumption they call `Locality'. 

A deep objection to this EPR-Schr\"odinger argument emerged in the work of John Stewart Bell in the 1960s \cite{Bell64,Myrvold21}. Bell considers similar experiments on pairs of entangled particles (specifically, a variation due to David Bohm \cite{Bohm51}). Referring to EPR's viewpoint, he summarises his own argument like this:
\begin{quote}
In this note that idea [of EPR's] will be \ldots\ shown to be incompatible with the statistical predictions of quantum mechanics. It is
the requirement of \textit{locality,} or more precisely \textit{that the result of a measurement on one system be unaffected
by operations on a distant system with which it has interacted in the past,} that creates the essential difficulty. \cite[195, emphasis added]{Bell64}
\end{quote}
The relevant `statistical predictions of quantum mechanics', now often called `Bell correlations', have since been confirmed many times. In 2022 the Nobel Prize for Physics was awarded to Alain
Aspect, John Clauser, and Anton Zeilinger, for some of the most important of these experiments. From an experimental viewpoint,  then, `entanglement is confirmed in its strangest
aspects', as Aspect put it in his
\href{https://www.nobelprize.org/prizes/physics/2022/aspect/speech/}{{speech}}
at the Nobel Prize banquet.

However, the conceptual implications of these results remain unclear, and vigorously contested. One central difficulty is the fact that nonlocality appears to be in tension with special relativity, though the existence and precise nature of the tension (if any), is itself a matter for dispute. And there is no agreed \textit{mechanism} for entanglement, if by mechanism we mean a category of physical relation, characterised in more general terms, that explains the nature of this new connection found in the quantum world. 

Roger Penrose puts the latter point like this. Saying that in his view, `there are two quite distinct mysteries presented by quantum entanglement', he continues:\begin{quote}
The first mystery is the phenomenon itself. How are we to come to terms with quantum entanglement and \textit{to make sense of it in terms of ideas that we can comprehend,} so that we can manage to accept it as something that forms an important part of the workings of our actual universe? \cite[591, emphasis added]{Penrose04}   
\end{quote} 
In this piece we propose a mechanism for entanglement, in this sense: a way of making sense of it in terms that are comparatively easy to comprehend.\footnote{We have nothing to offer here concerning Penrose's second mystery -- `[W]hy is it something that we barely notice in our direct experience of the world? Why do [the] \ldots\ effects of entanglement not confront us at every turn?' \cite[591]{Penrose04} -- though we will return briefly to these and related questions at the end of the paper. \label{fn:penrose}}  If correct, the proposal not only throws new light on this central feature of QM, but also, interestingly, seems to resolve the apparent tension with relativity. At any rate, it shows how Bell correlations can arise from components that are not themselves in tension with relativity.

We begin with a brief survey of the landscape of discussions of the implications of Bell's Theorem, in order to explain where our proposal sits in relation to other approaches.

 \section{EPR, Schr\"odinger, Bell, and Reichenbach}
In the field of views that descend from the work of EPR, Schr\"odinger and Bell, a helpful landmark is Reichenbach's Principle of the Common Cause (PCC) \cite{Reich56}. The following formulation of PCC  will do for our purposes \cite{Hofer2013}: 
\begin{quote}
The Common Cause Principle says that every correlation is either due to a direct
causal effect linking the correlated entities, or is brought about by a third factor,
a so-called common cause.
\end{quote}
Interpreted with reference to PCC, the relevant history goes like this. In 1935 EPR noted that correlations implied by what Schr\"odinger soon dubbed entanglement seemed to require explanation by common causes, not present within QM itself. EPR concluded that QM was incomplete, and Schr\"odinger agreed. The alternative -- the other option allowed by PCC -- was that measurement choices on one side of an experiment could influence results on the other, even though the two sides might be arbitrarily far apart, and indeed spacelike separated, in the sense of special relativity.\footnote{In other words, neither spacetime location is accessible from the other, by a signal limited by the speed of light.} To EPR and Schr\"odinger, that sort of action at a distance seemed absurd.\footnote{Their objection does not rest primarily on relativity, but on more general concerns about action at a distance. It is strengthened by the tension with relativity, however.} 

In the 1960s, however, Bell proved that under plausible assumptions, the QM predictions for some such experiments imply that the common cause option is untenable. That seems to leave us, as the above formulation of PCC puts it, with `a direct causal effect linking the correlated entities'; and hence with the kind of nonlocality that EPR and  Schr\"odinger had dismissed. As Bell saw, of course, this meant at least a \textit{prima facie} conflict with relativity.\footnote{In particular, it seems to require a preferred inertial frame in which this nonlocal action is instantaneous. Bell  concludes his 1964 paper as follows:  
\begin{quote}
In a theory in which parameters are added to quantum mechanics to determine the results of individual measurements, without changing the statistical predictions, there must be a mechanism whereby the setting of one measuring device can influence the reading of another instrument, however remote. Moreover, \textit{the signal involved must propagate instantaneously,} so that such a theory could not be Lorentz invariant. \cite[199, emphasis added]{Bell64}   
\end{quote}} 

In broad-brush terms, we can classify responses to Bell's argument as follows. This taxonomy is not comprehensive or precise, but it will serve to locate our current proposal.\footnote{We set aside challenges to the claimed experimental verification of Bell correlations.} 
\begin{enumerate}
    \item \textbf{Accept nonlocality}, acknowledging its conflict with relativity. Most such responses seek to mitigate the conflict, by arguing, for example, that Bell nonlocality is not the kind of full-blown (signalling) causation that would be in serious conflict with relativity; or that the preferred frame required by nonlocality is not empirically detectable.\footnote{See \cite{Maud11} for a comprehensive defence of this option.}

    \item \textbf{Avoid nonlocality}, by arguing that Bell's result depends on an assumption of `Realism' or `Classicality', and rejecting this assumption.\footnote{It is a common view that confirmation of the Bell correlations excludes \textit{local realism,} but leave us with two options: reject \textit{locality} or reject \textit{realism}. See \cite{Maud14,Gomori23} for critical discussion, and \cite{Wise17} for a sympathetic treatment.}  
 
        \item \textbf{Render nonlocality compatible with relativity}, by making it an \textit{indirect,} partially \textit{retrocausal} process, acting via the past light cones of the two observers. This requires that we abandon an assumption of Bell's argument called \textit{Statistical Independence} (SI), to allow measurement choices to influence hidden variables (HVs) in their past.\footnote{See \cite{WhartonArgaman20,FriedrichEvans19} for reviews of this approach.} Note that this option requires a choice about  the term `nonlocality'. There is a narrow use of the term, implying \textit{direct} spacelike influence, as in option (1), and a broad use, allowing the indirect influence proposed here. Those who prefer the narrow use will regard this option as another way of \textit{avoiding} nonlocality. Either way, this option agrees with option (1) that, as Bell puts it, `the result of a measurement on one system [is not] unaffected by operations on a distant system with which it has interacted in the past.' It thus falls into the same category as option (1) with respect to PCC.
    \item \textbf{Restore common causes} (and so avoid nonlocality altogether), by treating measurement settings as variables influenced by factors in the common past of the experimenters. This option, called \textit{superdeterminism}, also requires violation of SI, and  hence is sometimes confused with option (3).\footnote{Some  authors, though not themselves guilty of this confusion, use the term `superdeterminism' for any view rejecting SI \cite{Hoss19}.} As this classification shows, however, it rests on a different choice between the two alternatives offered by PCC.
    \item \textbf{Seek to avoid the problem}, by arguing that the Bell correlations are not subject to PCC. Here there are at least two previous proposals. 
    \begin{enumerate}
        \item Arguing that the Bell correlations arise from the fact that our viewpoint as observers is always `perspectival', e.g., confined to one branch of a larger set of ‘many worlds’ (with no need
for nonlocality in the bigger picture).\footnote{As \cite{Myrvold21} put it, Bell's `entire analysis is predicated on the assumption that, of the potential outcomes of a given experiment, one and only one occurs, and hence that it makes sense to speak of \textit{the} outcome of an experiment.' }
    \item Rejecting PCC altogether, arguing that the lesson of the Bell correlations is simply that the world contains robust patterns of correlations not explicable in the two ways that PCC allows.\footnote{See \cite{vanF82} for a view of this kind. This view does not avoid nonlocality, of course. It merely declines to explain it as PCC requires.}
    \end{enumerate}  
\end{enumerate}
\subsection{Our approach}

Our proposal has points in common with a number of these options. It agrees with (1) and (3) that measurement choices on one system may {make a difference} to the results of a measurement on a distant system, and with (3) in particular that this fact can be rendered compatible with relativity. 

We use the phrase `make a difference' for two reasons. First, it seems sufficiently neutral to avoid the terminological difficulties with the term `nonlocality', noted in (3) above.  Second, even if nonlocality is agreed to be real in the sense of option (1), its precise characterisation  is by no means straightforward. It is contentious, for example, whether such influences  are really \textit{causal}.\footnote{For discussion of such issues, see, e.g.:~\cite{Maud11,Wood15,Wise17,Myrvold21}.} We will not discuss this issue head-on, because our main proposal brings a new option to the table. However, it will be convenient to assume from the beginning that the relation in question is counterfactual-supporting, in the sense naturally conveyed by saying that measurement choices on one side of an Bell experiment \textit{make a difference} to outcomes on the other side, in some cases.\footnote{We stress that taking \textit{making a difference} to be a counterfactual-supporting notion need not imply the possibility of signalling. Making a difference can be entirely uncontrollable for signalling purposes. Example: I toss a fair coin twice, allowing you to choose \textit{beforehand} which of the two outcomes will be recorded. Your choice often \textit{makes a difference} to the recorded result, but you can't use that influence to embed a signal in the list, no matter for how long we play.}

Coming back to comparisons, our approach also agrees with option (5), in arguing that Bell correlations fall outside the scope of PCC. However, it locates them in what, QM aside, is a very familiar part of the landscape. In the causal modelling literature, it is a familiar fact that correlations may arise by `conditioning on a collider', and that such correlations are not subject to PCC. Our proposal is that Bell correlations are a (special sort of) correlation of this kind. We next introduce these ideas in general terms, before returning to the QM case in \S4.

\section{Conditioning on a collider}

In the language of causal models, a \textit{collider} is a variable with more than one direct cause, within a causal model. In other words, in the graphical format of directed acyclic graphs (DAGs), it is a node at which two or more arrows converge (hence the term `collider'). 

It is well known that `conditioning' on such a variable -- i.e., selecting the cases in which it takes a certain value -- may induce a correlation between its  causes, even if they are actually independent. As \cite[417]{Cole10} put it, `conditioning on the common effect imparts an association between two otherwise independent variables; we call this selection bias.'\footnote{Collider bias is also called {Berkson's bias,} after a Mayo Clinic physician and
statistician who called attention to it in the 1940s \cite{Berkson46}. But the point dates back at least to the Cambridge economist A C Pigou \cite{Pigou11}. (We are grateful to Jason Grossman and George Davey Smith here.)}

We'll give two simple examples of collider bias. For the first,\footnote{For which we are again indebted to George Davey Smith.} suppose that Ivy College selects students for academic or athletic excellence, not requiring both, and that these attributes are independent in the general population. 
The admission policy implies that if we learn that an Ivy student is not athletic we can infer that they are academically talented, and vice versa. Within this population, then,  there is a strong anti-correlation between the two attributes. But it is a selection artefact, a manifestation of collider bias. Admission to Ivy is a collider variable, influenced both by athletic ability and academic ability. The correlation arises because the selection rule sets aside all individuals who are neither athletically nor academically talented.

This example illustrates an important point. \textit{Correlations of this kind do not support counterfactuals.} Suppose that Holly is an Ivy track star, who struggles to get As. Would she have had better grades if she'd been less athletic? No. If she had been less athletic she wouldn't have been admitted to Ivy in the first place. (We will say that such correlations are \textit{counterfactually fragile.})

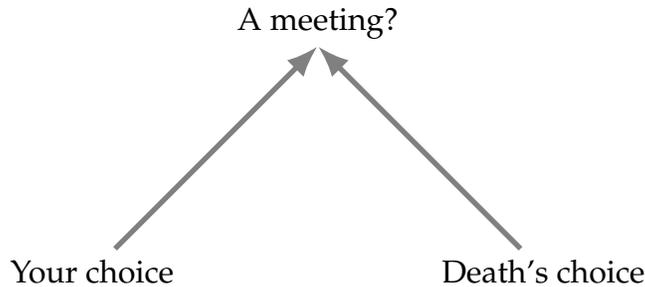
\begin{figure}
\centering
\begin{tikzpicture}
    \node (a) at (0,0) {Your choice};
    \node (F) at (3,3.3) {A meeting?};
    \node (b) at (6,0) {Death's choice};
\coordinate (coll) at (3,3); 
  \path[color=gray,rounded corners,line width=2pt] (a) edge (coll);
   \path[color=gray,rounded corners,line width=2pt] (b) edge (coll);

\end{tikzpicture}
\caption{A simple collider (`Death in Damascus')} \label{fig:M1}
\end{figure}

Our second example is the Death in Damascus case, well-known in the philosophy of  causation and decision theory \cite{Gibbard78}. Suppose that you and Death are each deciding where to travel tomorrow. You have the same two possible destinations -- in the usual version, Damascus and Aleppo. As in Figure~\ref{fig:M1}, your choice and Death's choice both influence an (aptly named) collider variable, which determines your survival. Let this variable take value $0$ if you and Death do not meet, and $1$ if you do. (We assume for simplicity that if the two of you choose the same destination, you will meet; and that this will be fatal, from your perspective.)

If we sample statistics for this kind of case, across a large population, they may suggest that people have an uncanny ability to evade Death, always choosing the opposite destination. If so, it is because we have introduced selection bias, by interviewing only survivors; such cases are sometimes called `survivorship bias' \cite{Czeisler21}.  

If you are a survivor, you might think to yourself, `I'm a survivor, so if I had chosen the other destination, Death would also have made a different choice.' You would be wrong. If you had chosen the other destination, you wouldn't have been a survivor. Again, this illustrates the fact that correlations resulting from conditioning on a collider do not support counterfactuals (in normal circumstances -- we'll come to an exception in a moment).

 \subsection{Constrained colliders}

Colliders are well-known in causal modelling, but our proposal requires an unusual modification of the familiar notion: namely, what we have called a \textit{constrained} collider \cite{PriceWharton21b,PriceWharton22}. Intuitively, this amounts to a restriction imposed from outside a causal model, biasing or completely specifying the value of the variable at the collider in question. 

In Death in Damascus, for example, we saw that your choice and Death's both influence a collider variable taking value $0$ if you do not meet, and $1$ if you do. If Fate wants to ensure that your number's up, as it were, she \textit{constrains} this collider, setting its value to $1$. In effect, Fate imposes a future boundary condition, \textit{requiring} that the collider variable take value $1$.

This boundary condition makes a big difference to the counterfactuals. If it weren't for Fate's role, your grieving relatives would be entitled to say, `If only you had made the other choice, you would still be with us today!' Once Fate constrains the collider, this is no longer true. If you'd made the other choice you would have met Death in the other place, instead. 

With Fate constraining the collider, in other words, there is a counter-factual-supporting connection between your movement and Death's. As \cite{PriceWeslake10} point out, you control Death's movements, in effect.\footnote{Price and Weslake are discussing the proposal that the ordinary past-to-future direction of causation rests on a low entropy boundary condition in the past -- on the so-called Past Hypothesis \cite{PriceWeslake10}. They use the Death in Damascus case to point out that a seemingly analogous future boundary constraint might produce not future-to-past causation, but the kind of zigzag influence exemplified in this Fate-constrained version of Death in Damascus. They suggest that a boundary constraint in the past might be expected to produce zigzags via the past, but don't connect this suggestion to quantum entanglement, as we shall do below.}  In \cite{PriceWharton21b,PriceWharton22}  we call this \textit{Connection across a Constrained Collider} ({CCC}) -- the terminology is deliberately non-committal about whether we regard it as causation, strictly speaking.

Collider constraint can come by degrees. Fate might be kind to you, and give you some small chance of eluding Death, at least tomorrow. We will be interested here in the full constraint version, in which case we'll say that the variable at the collider is \textit{locked.} 

A locked variable can only take one value -- that's the point. This means that it is no longer really a \textit{variable} at all, in the usual sense of a causal model, and can no longer be an effect of any of the remaining variables. We could put it like this: causation requires making a difference, and locking prevents making a difference. By locking the collider variable in the Death in Damascus case, Fate makes it the case that your choice makes no difference to whether you encounter Death. You no longer have any causal influence on the matter.

It is easy to invoke Fate, this time in a more friendly manner, to produce a constrained version of our Ivy College admissions example. If Fate has determined that Holly will be admitted to Ivy, one way or another, then if she hadn't been athletic she would have been academic.

These cases are highly unrealistic, of course. This doesn't matter here, where their point is simply to introduce the notion of a constrained collider. Later, when it matters, we'll rely on realistic constraints. However, we can now describe our goal, and our route for getting there.

\subsection{Outline of the proposal and argument}

Our proposal is that constrained colliders are the source of Bell correlations -- that entanglement itself is CCC (Connection across a Constrained Collider). Our argument for this conclusion will be in three steps. 
\begin{enumerate}
    \item  We will call attention to the fact that \textit{some} manifestations of Bell correlations, closely related to some of the leading experiments, have already been recognised to involve collider bias. In these particular cases, Bell correlations are known to be selection artefacts, and don't support the counterfactuals characteristic of nonlocality. 
   \item  We will then focus on a special subclass of these cases. We will point out that in these special cases, a proposal made in the physics literature for other purposes has the effect of \textit{constraining} the relevant colliders. This converts mere selection artefacts into CCC, which does support the relevant counterfactuals. In these unusual cases, then, we will have shown how robust, counterfactual-supporting Bell nonlocality can result from a combination of two things: (i) collider bias, and (ii) a boundary constraint, imposing a particular value on a collider variable. 
   \item We will then argue that the same explanation works in ordinary cases of entanglement, without the support of the proposal needed in (2) -- its role will be played by something much more familiar.
\end{enumerate}

\section{Entanglement swapping and delayed-choice}

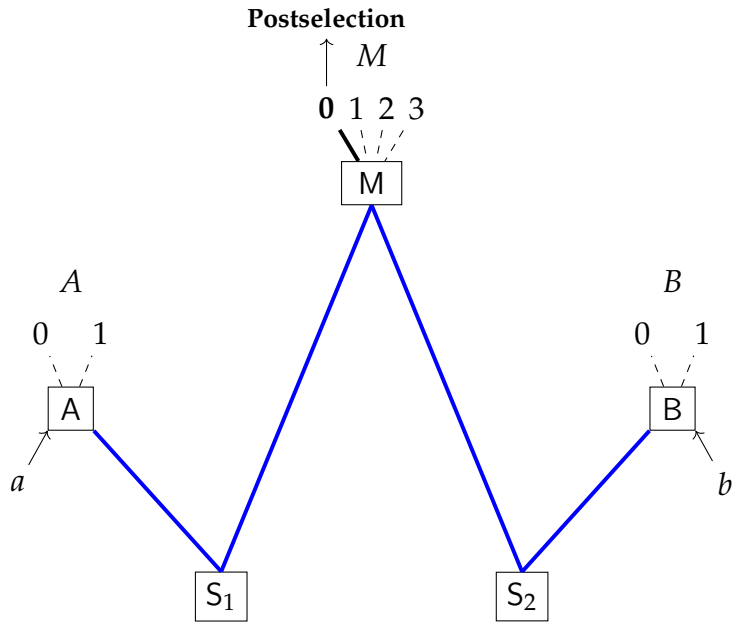
\begin{figure}[t]
\centering
\begin{tikzpicture}
    \node[draw,rectangle,minimum width=0.6cm,minimum height=0.5cm] (I1) at (3,-2.5) {\textsf{S$_1$}};
    \node[draw,rectangle,minimum width=0.6cm,minimum height=0.5cm] (I2) at (7,-2.5) {\textsf{S$_2$}};
    \node[draw,rectangle,minimum width=0.8cm,minimum height=0.5cm] (W) at (5,3) {\textsf{M}};
\node[] (W0) at (4.4,4) {\textbf{0}};
\node[] (W1) at (4.8,4) {1};
 \node[] (W2) at (5.2,4) {2};
 \node[] (W3) at (5.6,4) {3};
\node[] (Out) at (5,4.7) {$M$};

 \path [dashed,-] (W1) edge (W);
 \path [line width=1.5pt,-] (W0) edge (W);
\path [dashed,-] (W2) edge (W);
 \path [dashed,-] (W3) edge (W);

    \node[draw,rectangle,minimum width=0.6cm,minimum height=0.5cm] (Abox) at (1,0) {\textsf{A}};
  \node[] (Alabel) at (1,1.7) {$A$};
    \node[] (A0) at (0.6,1) {0};
    \node[] (A1) at (1.4,1) {1};
    \node[] (a) at (0.3,-1) {$a$};
    \node[draw,rectangle,minimum width=0.6cm,minimum height=0.5cm] (Bbox) at (9,0) {\textsf{B}};
      \node[] (Blabel) at (9,1.7) {$B$};
       \node[] (B0) at (8.6,1) {0};
    \node[] (B1) at (9.4,1) {1};
     \node[] (b) at (9.7,-1) {$b$};
       \node[] (bc) at (4.4,5.2) {\footnotesize{\textbf{Postselection}}};
       \path [<-] (bc) edge (W0);

   \path [dashed,-] (Abox) edge (A0);
   \path [dashed,-] (Abox) edge (A1);
    \path [->] (a) edge (Abox.south west);
    \path [dashed,-] (Bbox) edge (B0);
   \path [dashed,-] (Bbox) edge (B1);
   \path [->] (b) edge (Bbox.south east);
      \draw [color=blue,line width=1.5pt,-] (I1.north) -- (Abox.south east);
   \draw [color=blue,line width=1.5pt,-] (I2.north) -- (Bbox.south west);
      \draw [color=blue,line width=1.5pt,-] (I1.north) -- (W.south);
       \draw [color=blue,line width=1.5pt,-] (I2.north) -- (W.south);

  
\end{tikzpicture}
\caption{The W-shaped case with postselection for outcome $0$ at \textsf{M}} \label{fig:W}
\end{figure}

To make a start, let's return to the quantum cases. Figure~\ref{fig:W} depicts an important experimental protocol, used for some of the best tests of Bell correlations. We call it the W-shaped protocol, referring to the shape of its arrangement in spacetime. It relies on so-called \textit{entanglement swapping.} Two pairs of entanglement particles are produced independently, one pair at the source \textsf{S$_1$} and the other at the source \textsf{S$_2$}. One particle from each pair is sent to a measurement device at \textsf{M}, which performs a particular joint measurement on the two particles -- a measurement that has four possible outcomes, shown in Figure~\ref{fig:W} as $0,1,2$ and $3$.

If we `postselect' the cases in which the measurement \textsf{M} takes a particular value\footnote{That is, in effect, if we discard all the other cases.} -- say, $0$, as shown in Figure~\ref{fig:W} -- then each remaining pair of particles exhibits the correlations typical of entanglement, \textit{even though the particles concerned have never interacted}. This is entanglement swapping, and it can be used to test for Bell correlations. Some of the most impressive recent results, the so-called `loophole-free' Bell experiments, use this protocol.\footnote{See \cite{Hensen15,Giustina15,Shalm15}.} 

This is what is shown on the two wings of Figure~\ref{fig:W}. Each of the measuring devices \textsf{A} and \textsf{B} has two possible settings, chosen by the experimenters, Alice and Bob, at those locations, and two possible outcomes. Each run of the experiment thus produces four binary numbers $\{a,b,A,B\}$, comprising two settings and two outcomes, with sixteen possible combinations. Bell correlations are facts about the relative frequencies of each of the sixteen possibilities, in long sequences of trials.

In principle, experiments of this form can be performed in which the measurement \textsf{M} has any one of three different temporal locations, with respect to those at \textsf{A} and \textsf{B}. \textsf{M} can be in the past (i.e., in the overlap of the past light cones of \textsf{A} and \textsf{B}), in the future (i.e., the overlap of the future light cones), or spacelike separated from \textsf{A} and \textsf{B}. (There are also some mixed cases, which we can ignore here.) In the three recent experiments cited above, two used the version in which \textsf{M} was in the past with respect to \textsf{A} and \textsf{B}, and one used spacelike separation. 

The case that interests us, and the one actually depicted in Figure~\ref{fig:W}, is the remaining possibility, in which \textsf{M} occurs strictly \textit{later} than \textsf{A} and \textsf{B}. This case relies on \textit{delayed-choice} entanglement swapping (DCES). So far as we are aware, no loophole-free Bell experiment has been performed in this version, although the experimental predictions are the same as in the other two cases (and other kinds of delayed-choice entanglement swapping measurements have been performed \cite{Ma12}, with successful results). 

Several authors have noted that in the delayed-choice version of the W-shaped experiment, Bell correlations in the postselected ensemble may be selection artefacts, a manifestation of collider bias.\footnote{See \cite{Gaasbeek10,Egg13,Fankhauser19,Guido21,PriceWharton21a, Mjelva24}.}  A mark of this is that these correlations need not support the counterfactuals we are taking to be characteristic of Bell nonlocality. A change in Alice's setting $a$ may produce a change in the result $M$ of the intermediate measurement \textsf{M}, thus removing the case from the postselected ensemble. If so, there is no need  for it to make a difference to Bob's outcome $B$. We may think of the options at \textsf{M} as a kind of \textit{safety valve,} capable of releasing the pressure for change arising from changes in the settings at \textsf{A} and \textsf{B}. In Bell experiments of this kind, then, the Bell correlations themselves may be counterfactually fragile.\footnote{In previous work, we call this the \textit{Collider Loophole} for W-shaped tests of nonlocality, and discussed which W-shaped Bell tests are subject to it \cite{PriceWharton21a}. For present purposes, we need only the DCES case, where it is uncontroversial.}

Two notes about these cases. First, it is one thing to point out that these cases \textit{may} involve collider bias, another to show that this is actually true in practice. The second step requires a causal mechanism whereby the settings choices at \textsf{A} and \textsf{B} influence the outcome at \textsf{M}, and the details of that are likely to depend on one's assumed quantum ontology. But a recent paper by Mjelva works through some of these details \cite{Mjelva24}.

Second, an objection. Since our professed goal is to explain Bell correlations in relativity-friendly terms, it seems relevant that the intuitive causal story in the DCES W case (Figures~\ref{fig:W} and \ref{fig:W2}) is \textit{not} confined to lightcones. In standard QM, for example, it involves a direct, `instantaneous' projection on the inner particles, within each wing of the experiment, when measurements are performed at \textsf{A} and \textsf{B}. So, even if we explain away the apparent action at a distance between \textsf{A} and \textsf{B}, we'll still be relying on it, apparently, within each V-shaped wing of the W. 

This objection is correct as far as it goes, but as we will see, it does not have any bearing on the ability of the proposal to provide a relativity-friendly explanation of nonlocality in the V-shaped case. And once it is shown to work in V-shaped cases, it can be applied in particular to the V-shaped wings of the W experiment, eliminating the need for instantaneous projection.

\subsection{A toy example}

Some readers may find it surprising that Bell correlations can arise as postselection artefacts. However, it is easy to construct simple models with the same feature. Such toy models don't explain how QM achieves the result, but they do show that the trick itself is not difficult, once postselection enters the picture. 

Here's an example. Suppose that Alice and Bob each choose a `setting', $0$ or $1$, and also produce an `outcome' by tossing a fair coin -- let's use $0$ and $1$ for those results, too. Each then sends both numbers to a third observer, Charlie. At each trial, Charlie thus receives four binary numbers $\{a,b,A,B\}$, just as in the kind of Bell experiments we are discussing. It is very easy for Charlie to sort these results into four bins, corresponding to the four possible outcomes $M$ of the central measurement \textsf{M} in Figures~\ref{fig:W}, so that the statistics within each bin are identical to those in the corresponding subensemble of that experiment. If Charlie then `postselects', by throwing away the results in three of the bins, the results in the remaining bin display Bell correlations, exactly as in the corresponding selection in the real experiment. 

Taken at face value, these correlations might suggest that the results of the coin tosses on one side of the experiment are not unaffected by settings choices on the other. (We can build it in to the model that there are no common causes, influencing both sides of the experiment.) But it is clear that this a selection artefact. Alice's setting choices certainly make a difference to which bin Charlie selects for the result, in some cases; but no difference at all to Bob's coin toss. Just as in Figures~\ref{fig:W}, the choice of bin acts as a safety valve, absorbing the effects of changes to either setting.

\subsection{Constraining a quantum collider}

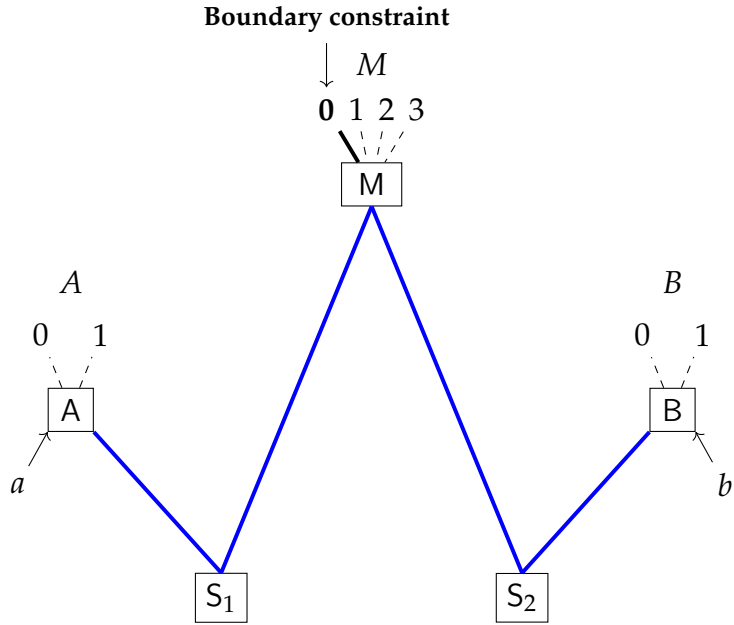
\begin{figure}[t]
\centering
\begin{tikzpicture}
 \node[draw,rectangle,minimum width=0.6cm,minimum height=0.5cm] (I1) at (3,-2.5) {\textsf{S$_1$}};
    \node[draw,rectangle,minimum width=0.6cm,minimum height=0.5cm] (I2) at (7,-2.5) {\textsf{S$_2$}};
    \node[draw,rectangle,minimum width=0.8cm,minimum height=0.5cm] (W) at (5,3) {\textsf{M}};
\node[] (W0) at (4.4,4) {\textbf{0}};
\node[] (W1) at (4.8,4) {1};
 \node[] (W2) at (5.2,4) {2};
 \node[] (W3) at (5.6,4) {3};
  \node[] (Out) at (5,4.6) {$M$};
   \node[] (bc) at (4.4,5.2) {\footnotesize{\textbf{Boundary constraint}}};


 \path [dashed,-] (W1) edge (W);
 \path [line width=1.5pt,-] (W0) edge (W);
\path [dashed,-] (W2) edge (W);
 \path [dashed,-] (W3) edge (W);
\path [->] (bc) edge (W0);

    \node[draw,rectangle,minimum width=0.6cm,minimum height=0.5cm] (Abox) at (1,0) {\textsf{A}};
  \node[] (Alabel) at (1,1.7) {$A$};
    \node[] (A0) at (0.6,1) {0};
    \node[] (A1) at (1.4,1) {1};
    \node[] (a) at (0.3,-1) {$a$};
    \node[draw,rectangle,minimum width=0.6cm,minimum height=0.5cm] (Bbox) at (9,0) {\textsf{B}};
      \node[] (Blabel) at (9,1.7) {$B$};
       \node[] (B0) at (8.6,1) {0};
    \node[] (B1) at (9.4,1) {1};
     \node[] (b) at (9.7,-1) {$b$};

   \path [dashed,-] (Abox) edge (A0);
   \path [dashed,-] (Abox) edge (A1);
    \path [->] (a) edge (Abox.south west);
    \path [dashed,-] (Bbox) edge (B0);
   \path [dashed,-] (Bbox) edge (B1);
   \path [->] (b) edge (Bbox.south east);

      \draw [color=blue,line width=1.5pt,-] (I1.north) -- (Abox.south east);
   \draw [color=blue,line width=1.5pt,-] (I2.north) -- (Bbox.south west);
      \draw [color=blue,line width=1.5pt,-] (I1.north) -- (W.south);
       \draw [color=blue,line width=1.5pt,-] (I2.north) -- (W.south);


\end{tikzpicture}
\caption{The W-shaped case with outcome $0$ at \textsf{M} imposed by the Horowitz-Maldacena boundary constraint}\label{fig:W2}
\end{figure}

In DCES W-shaped Bell tests, then, Bell correlations are plausibly explained as selection artefacts, rather than manifestations of true, counterfactual-supporting nonlocality. However, we now wish to consider the case depicted in Figure~\ref{fig:W2}, in which the safety valve at \textsf{M} is closed, because the measurement there is constrained to take a particular value.

It turns out that this case can be found in the physics literature,  
in the form of a proposal from Horowitz and Maldecena concerning the   black hole information paradox \cite{HorowitzMaldacena04}. 
The background here is that Stephen Hawking discovered a process now called Hawking radiation, by which black holes eventually evaporate away to nothing. He thought initially that this process would be random, preventing the escape of information that had fallen into the black hole in the first place, and hence in conflict with the usual reversibility, or `unitarity', of quantum theory. Various mechanisms to allow escape of this information were then proposed, including the one that interests us here.

Horowitz and Maldacena describe their proposal as follows:  
\begin{quote}
In [Hawking's] process of black hole evaporation, particles are created in correlated pairs with one
falling into the black hole and the other radiated to infinity. The correlations remain even
when the particles are widely separated. The final state boundary condition at the black
hole singularity acts like a measurement that collapses the state into one associated with
the infalling matter. This transfers the information to the outgoing Hawking radiation in
a process similar to ``quantum teleportation''. \cite{HorowitzMaldacena04}
\end{quote}

In other words, the process of Hawking radiation creates pairs of entangled particles. One member of each pair, $X_{in}$, falls inside the  the black hole; the other member, $X_{out}$, escapes to the outside world. When a different particle $Y$ falls into a black hole, it eventually encounters a final boundary condition at the singularity, meeting up with $X_{in}$.  As Horowitz and Maldacena say, this boundary condition `acts like a measurement', but without the usual safety valve. As in Figure~\ref{fig:W2}, the boundary condition only allows one possible measurement result. 
Because $X_{in}$ is entangled with $X_{out}$, the boundary condition at the singularity forces particle $X_{out}$ to be a precise quantum copy of the original in-falling particle $Y$, in the same sense as copies produced by quantum teleportation.

Any proposal to recover complete information from a black hole also has to allow for an even more problematic case: when in place of the single particle $Y$ there are two entangled particles $Y_{in}$ and  $Y_{out}$, only one of which falls into the black hole. 
 In order to resolve the information loss paradox in this case, the original entangled state of $Y_{in}$ and  $Y_{out}$ would somehow have to be restored.  
 
 The procedure for accomplishing this in the Horowitz-Maldacena framework looks essentially like Figure~\ref{fig:W2}.  Here $Y_{in}$ and  $Y_{out}$ are created at $S_1$, where the particle on the right ($Y_{in}$) falls into a black hole.  Meanwhile, the Hawking pair is created at $S_2$, with the left particle ($X_{in}$) heading into the black hole.  These two particles meet at the future singularity ($M$), and the constrained measurement leads to the conclusion that the two particles outside the black hole, on the sides of the $W$ -- i.e., $Y_{out}$ and $X_{out}$ -- must be in the same entangled state as originally true of $Y_{in}$ and  $Y_{out}$ at $S_1$.  

To resolve the information paradox, this process must happen with certainty, not merely probabilistically. So the final state boundary condition at the singularity inside the black hole needs to impose a particular result on the measurement, eliminating the usual need for postselection.  In our terminology, this amounts to constraining a collider at that point.

Discussing the Horowitz-Maldacena hypothesis recently, Malcolm Perry puts it like this:

\begin{quote}
{[}t{]}he interior of the black hole is therefore a strange place where
one's classical notions of causality \ldots{} are violated. This does
not matter as long as outside the black hole such pathologies do not
bother us. \cite[9]{Perry21}
\end{quote}
As we'll explain, our proposal is going to be that boundary conditions doing this job are
actually extremely common, if you know where to look. In the other
direction of time, they are just familiar, controllable constraints on the initial conditions of experiments, and don't need black holes.

For the moment, the point we need is that the Horowitz-Maldacena proposal does for measurement inside a black hole what Fate does in our two examples earlier. It \textit{locks} a collider variable -- in this case, the final measurement outcome -- to a particular value. Maldacena and Horowitz propose that this creates a zigzag causal path, along which information can escape from a black hole. It is crucial to their proposal that this process supports counterfactuals. If it is to explain how information escapes from black holes, then it needs to be counterfactually robust: if different information had fallen into the black hole in the first place, different information would have emerged. 

\begin{figure}[t]
\centering
\begin{tikzpicture}
    \node[draw,rectangle,minimum width=0.6cm,minimum height=0.5cm] (I) at (3.1,-2.5) {\textsf{C}};
\node[] (I0) at (2.5,-3.4) {\textbf{0}};
 \node[] (I1) at (2.9,-3.4) {1};
 \node[] (I2) at (3.3,-3.4) {2};
 \node[] (I3) at (3.7,-3.4) {3};
 \node[] (Prep) at (2.5,-4.7) {\textbf{Preparation}};
 \node[] (C) at (3.1,-4) {$C$};

  \path [dashed,-] (I1) edge (I);
  \path [line width=1.5pt,-] (I0) edge (I);
  \path [dashed,-] (I2) edge (I);
  \path [dashed,-] (I3) edge (I);
  \path [->] (Prep) edge (I0);

    \node[draw,rectangle,minimum width=0.6cm,minimum height=0.5cm] (Abox) at (1.2,0) {\textsf{A}};
   \node[] (Alabel) at (1.2,1.7) {$A$};
    \node[] (A0) at (0.8,1) {0};
    \node[] (A1) at (1.6,1) {1};
    \node[] (a) at (0.5,-1) {$a$};
    \node[draw,rectangle,minimum width=0.6cm,minimum height=0.5cm] (Bbox) at (5,0) {\textsf{B}};
     \node[] (Blabel) at (5,1.7) {$B$};
       \node[] (B0) at (4.6,1) {0};
    \node[] (B1) at (5.4,1) {1};
    \node[] (b) at (5.7,-1) {$b$};

   \path [dashed,-] (Abox) edge (A0);
   \path [dashed,-] (Abox) edge (A1);
   \path [->] (a) edge (Abox.south west);
    \path [dashed,-] (Bbox) edge (B0);
   \path [dashed,-] (Bbox) edge (B1);
    \path [->] (b) edge (Bbox.south east);

   \draw [color=blue,line width=1.5pt,-] (I.north) -- (Abox.south east);
   \draw [color=blue,line width=1.5pt,-] (I.north) -- (Bbox.south west);

\end{tikzpicture}
\caption{The V-shaped case with preparation in state \textbf{0} at input $C$} \label{fig:V}
\end{figure}
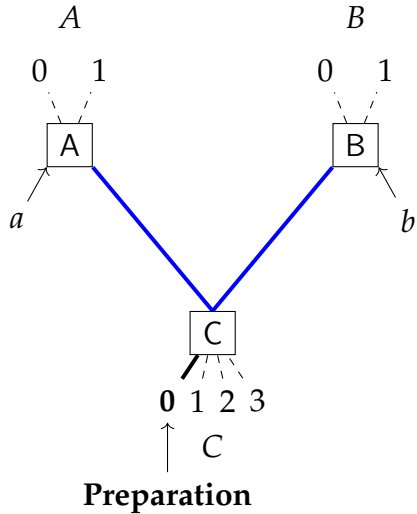

In this very special case, then, we have shown how robust counterfactual-supporting Bell nonlocality can result from a combination of two things: (i) collider bias, and (ii) a boundary constraint, imposing a particular value on a collider variable. The remaining step is to explain how this model fits much more ordinary cases of entanglement, such as the V-shaped Bell experiments shown in Figure~\ref{fig:V}.

\section{Application to V-shaped cases (I)}
Figure~\ref{fig:V} shows a typical Bell experiment, with the same kind of `V-shaped' two-particle geometry as the experiments discussed by EPR and Schr\"odinger in 1935. A pair of particles is prepared in a certain entangled state at \textsf{C}, where there are four options, again labelled $0,1,2$ and $3$; Figure~\ref{fig:V} shows the case in which this initial state is chosen to be $0$. These two entangled particles are sent to measurements at \textsf{A} and \textsf{B}, where the experimeters choose binary measurement settings $a$ and $b$, as before. The measurements \textsf{A} and \textsf{B} each produce an outcome, either $0$ or $1$. The four values $\{a,b,A,B\}$ are recorded, and (for appropriate versions of this experiment), QM predicts Bell correlations in the relative frequencies in series of such results. 

It is important that the initial state is held fixed, or at any rate not allowed to vary entirely at random. If we mix together results from the four possible versions of this experiment, randomly and in equal proportion, Bell correlations will no longer be visible in the results. In that sense, the experiment already relies on the kind of `preselection' that our ordinary control of the initial state makes possible.

 Readers who see where this observation might be leading will also see the obvious obstacle. True enough, Bell correlations require that we `preselect' on the input variable at \textsf{C} in Figure~\ref{fig:V}. But this variable does not seem to be a collider. In the familiar case, in which the measurement is simply \textit{prepared} in a particular initial state at \textsf{C}, that initial state is obviously not influenced by the measurement settings at \textsf{A} and \textsf{B}. 

However,  this is not the relevant case to consider. After all, the final state at \textsf{M} isn't affected by Alice and Bob's choice of measurement settings, in the case in which it is constrained, as in Figure~\ref{fig:W2}. In that case, the outcome at \textsf{M} is locked by the boundary constraint. What matters is that it's a collider \textit{when not constrained.}

So in the V-shaped case, too, we need to consider the unconstrained case. Unlike in the W-shaped case, however, the unconstrained V-shaped case is highly unusual, and can easily escape notice altogether. We are going to propose two ways to bring it into view.  But before that, lest readers take fright before the proposal is even hatched, we want to tuck it under a familiar wing.

\section{Conditioning on the Past Hypothesis}

You are walking across campus one windy day, and notice some scraps of paper, falling from a window. Picking up one scrap, you find some familiar words: ``There are more things in heaven and earth, Hora \ldots'' -- the printed words tail off at the torn edge of the paper. 

What should you expect if you pick up a nearby scrap, with a matching torn edge? Well, it depends which building it is. If it is the primate lab, and you can hear the sound of typewriters, you might well expect gibberish. If it is the library, then more \textit{Hamlet,} or at any rate not gibberish.

The case illustrates one of the most important lessons of the physics of time-asymmetry, over the past 150 years. In the 1870s, Josef Loschmidt objected  that same statistical considerations that his colleague Ludwig Boltzmann took to explain why entropy should increase towards the future, should also imply that it increases towards the past. This would imply that the apparent order around us is, like the monkey's snippet of \textit{Hamlet,} a random piece of organisation in a stream of gibberish. These days, the point is often made by imagining so-called Boltzmann Brains -- random, self-aware thinkers, coalescing by chance from the background chaos. 

Loschmidt's challenge to Boltzmann is known as the reversibility objection. The objection itself, and the most influential present-day response to it, are here described by Mathias Frisch. 

\begin{quote}
If we assume an equiprobability distribution of micro-states compatible with a given
macro state of non-maximal entropy, then it can be made plausible that, intuitively, ‘most’ micro
states will evolve into states corresponding to macro states of higher entropy. However, if the
micro-dynamics governing the system is time-symmetric, then the same kind of considerations
also appear to show that, with overwhelming probability, the system evolved \textit{from} a state of
higher entropy. This undesirable retrodiction, which is at the core of the \textit{reversibility objection,}
can be blocked, if we \textbf{conditionalize} the distribution of micro-states not on the present macro-state but on a low-entropy initial state of the system. \ldots\ Albert and others argue that we are ultimately led to
postulate an extremely low-entropy state at or near the beginning of the universe. Albert  
calls
this postulate “the past hypothesis” \textit{(PH)}. \cite[15, emphasis in bold added]{Frisch07}    
\end{quote}

Thus, it is a common view that the familiar time-asymmetric thermodynamic character of our universe, and whatever depends on it, turns on a combination of two things:  (i) a time-symmetric probability distribution over the space of possible histories, or `trajectories', allowed by the relevant physical laws; and (ii) a time-asymmetric condition, the so-called Past Hypothesis (PH), specifying a low entropy macrostate for the universe shortly after the Big Bang. \textit{Conditionalizing} on PH simple means disregarding all the trajectories of which it is not true, at the relevant point in time. It is just like postselection, except that the possibilities we ignore are not \textit{actual,} but merely \textit{possible.} Note that this is the same as the difference between ordinary postselection, where we discard actual results, and the constrained version, illustrated by the Horowitz-Maldacena proposal, where the constraint excludes `merely possible' results -- i.e., cases that would be possible, were it not for the constraint in question.

 On the cosmological scale, the Past Hypothesis thus plays the role of the college library, in our little example above. It is the source of all the order we encounter in the universe\footnote{Including, of course, ourselves.}  --  the mother of all libraries, to adapt a phrase from \cite[118]{Albert02}.

\begin{figure}[t]
\centering
\includegraphics[width=12.5cm]{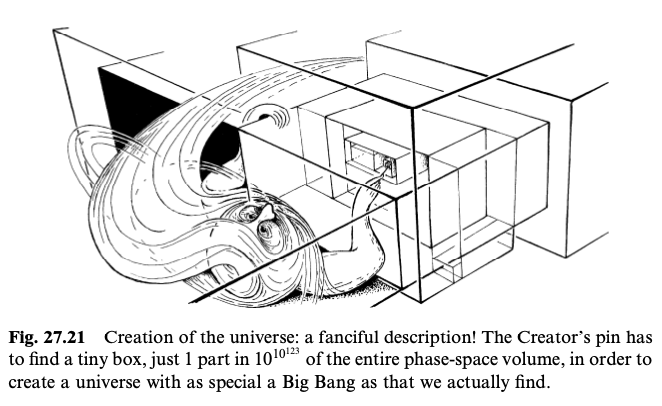}
\caption{Penrose on the Past Hypothesis} \label{fig:god}
\end{figure}

It is difficult to convey how big a restriction on the space of possible trajectories the Past Hypothesis needs to be.  \cite[730]{Penrose04}  has a go at it, calculating a figure of $1$ in $10^{{10}^{123}}$, and providing us with an illustration of the Creator choosing the initial state with a very, \textsc{very}, VERY fine pin. (See Figure~\ref{fig:god}.)

Summing up, our familiar universe seems to be massively `preconstrained', compared to a natural understanding of its space of possible histories, without PH. Our plan is to summon a bit of that preconstraint,  to explain how entanglement in ordinary V-shaped cases can be CCC.

\section{Application to V-shaped cases (II)}

Our goal is to bring the unconstrained V-shaped case into view. We propose two strategies. One of them is \textit{local,} in the sense that it focuses simply on experiments like that shown in Figure~\ref{fig:V}, and proposes a technique that might make the initial variable unconstrained \textit{in the real world.} Whether this is possible seems to be an open question, as we'll explain. The other strategy is \textit{global,} in that it asks us to consider whatever time-asymmetric feature of reality normally gives us control of the initial but not the final conditions of experiments -- plausibly, as we've just seen, this is the Past Hypothesis -- and to consider an idealised regime in which that feature is absent. In effect, then, the global strategy is to find the unconstrained V-shaped case in the huge set of universes that the Creator is discarding in Figure~\ref{fig:god}. As we'll explain, this will be sufficient for our main argument, even if the local strategy is unsuccessful; and is in any case necessary, because the local strategy gives us, at best, only a small subset of the cases we need.

\subsection{Local strategy: a delayed-choice V-shaped experiment?}

\begin{figure}[t]
\centering
\begin{tikzpicture}
    \node[draw,rectangle,minimum width=0.6cm,minimum height=0.5cm] (I) at (3.1,-2.5) {{\textsf{C}}};
\node[] (I0) at (2.5,-3.4) {0};
 \node[] (I1) at (2.9,-3.4) {1};
 \node[] (I2) at (3.3,-3.4) {2};
 \node[] (I3) at (3.7,-3.4) {3};
 \node[] (C) at (3.1,-4) {$C$};

 \path [dashed,-] (I1) edge (I);
  \path [dashed,-] (I0) edge (I);
  \path [dashed,-] (I2) edge (I);
  \path [dashed,-] (I3) edge (I);

 \node[] (D0) at (2.5,3.7) {0};
 \node[] (D1) at (2.9,3.7) {1};
 \node[] (D2) at (3.3,3.7) {2};
 \node[] (D3) at (3.7,3.7) {3};
 \node[] (Dout) at (3.1,4.3) {$D$};

     \node[draw,rectangle,minimum width=0.6cm,minimum height=0.5cm] (D) at (3.1,2.8) {\textsf{D}};

\path [dashed,-] (D1) edge (D);
  \path [dashed,-] (D0) edge (D);
 \path [dashed,-] (D2) edge (D);
 \path [dashed,-] (D3) edge (D);

    \node[draw,rectangle,minimum width=0.6cm,minimum height=0.5cm] (Abox) at (1.2,0) {{\textsf{A}}};
   \node[] (Alabel) at (1.2,1.7) {$A$};
    \node[] (A0) at (0.8,1) {0};
    \node[] (A1) at (1.6,1) {1};
    \node[] (a) at (0.5,-1) {$a$};
    \node[draw,rectangle,minimum width=0.6cm,minimum height=0.5cm] (Bbox) at (5,0) {{\textsf{B}}};
    \node[] (Blabel) at (5,1.7) {$B$};
       \node[] (B0) at (4.6,1) {0};
    \node[] (B1) at (5.4,1) {1};
    \node[] (b) at (5.7,-1) {$b$};

   \path [dashed,-] (Abox) edge (A0);
   \path [dashed,-] (Abox) edge (A1);
   \path [->] (a) edge (Abox.south west);
    \path [dashed,-] (Bbox) edge (B0);
   \path [dashed,-] (Bbox) edge (B1);
    \path [->] (b) edge (Bbox.south east);

   \draw [color=blue,line width=1.5pt,-] (I.north) -- (Abox.south east);
   \draw [color=blue,line width=1.5pt,-] (I.north) -- (Bbox.south west);


    \draw[color=red,line width=1.5pt] (a) to (Abox.south) to (I.north west);
    \draw[color=red,line width=1.5pt] (b) to (Bbox.south) to (I.north east);


     \draw[color=gray,dashed,rounded corners,line width=1pt] (I) -- (D);

\end{tikzpicture}
\caption{V-shaped case with delayed-choice measurement of initial state} \label{fig:V2a}
\end{figure}
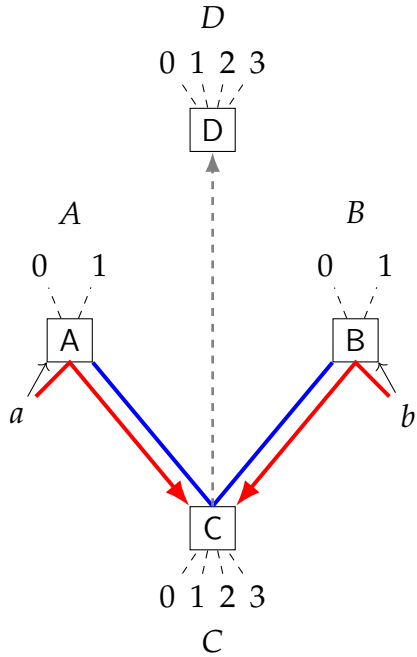

For the local argument, imagine the version of the V-shaped experiment shown in Figure~\ref{fig:V2a}, in which the four possible initial Bell states are produced at \textsf{C}, with equal probability. In each run of the experiment the initial state is recorded, but by a device that does not produce a classical record until a measurement \textsf{D}, in the absolute future of the measurements at \textsf{A} and \textsf{B}. (This is the delayed-choice element.)\footnote{We are indebted to Gerard Milburn for this proposal.}

The expected statistics for this experiment are straightforward. There are no Bell correlations in the statistics as a whole, but such correlations emerge if we postselect on any of the four possible results of \textsf{D} (or, equivalently, preselect on any of the four initial states at \textsf{C}).

How should we interpret these results? Are the Bell correlations in these postselected ensembles counterfactually \textit{fragile,} as in Figure~\ref{fig:W}, or counterfactually \textit{robust,} as in Figure~\ref{fig:W2}? The question turns on whether the settings choices at \textsf{A} and \textsf{B} are able to influence the results of the measurement at \textsf{D}. If such influence is possible, then we have our safety valve: the Bell correlations in each subensemble can again be explained as a selection artefact -- no counterfactually robust nonlocality is required. 

How should we answer this question? Indeed, does it have a settled answer, one way or the other, or is the case under-described? Perhaps there are different possible versions of such an experiment, some going one way and some the other. 

For our purposes, it isn't necessary to resolve these issues. For us, the case is a thought experiment. If the reader will allow the possibility of versions of this experiment in which the Bell correlations in each subensemble are a selection artefact, then we have the model we want. We have a V-shaped case with an unconstrained initial state, from which the V-shaped case in Figure~\ref{fig:V} may be obtained by reimposing ordinary measurement preparation.  In this transition, as in the W case, the Bell correlations are transformed from mere selection artefacts to counterfactual-supporting CCC.

However, if the setting choices at \textsf{A} and \textsf{B} make a difference to the outcome at \textsf{D}, then they also make a difference at \textsf{C}. The case then involves \textit{retrocausality,} as shown in red in Figure~\ref{fig:V2a}. 
Some readers will balk at this possibility. Invoking the words of John Wheeler, perhaps, they may object that it allows that `present choice influences past dynamics, in contravention of every principle of causality' \cite[41]{Wheeler78}.
It would take us too far afield to challenge Wheeler's instincts on this matter.\footnote{We have done so at length elsewhere: see \cite{Price96,PriceWharton15,PriceWharton16,WhartonArgaman20}.  In the present context, we may be overstating the relevance of Wheeler's objection. Wheeler is discussing a delayed-choice experiment, and the full passage runs like this:
\begin{quote}
    Does this result mean the present choice influences past dynamics, in contravention of every principle of causality? Or does it mean, calculate pedantically and don’t ask questions? Neither; the lesson presents itself rather as this, that the past has no existence except as it is recorded in the present.
\end{quote}
Wheeler's own option may well be sufficient for our purposes. If in Figure~\ref{fig:V2a} the measurement settings at \textsf{A} and \textsf{B} can make a difference to the past recorded at \textsf{D}, that seems good enough for our safety valve.} But for readers wedded to those instincts, we need a more powerful intuition pump. 

In any case, this local strategy is going to work, at best, only in a very restricted class of examples -- at most, cases in which it is not simply \textit{obvious} that the initial state is fixed, and hence not amenable to influence by later measurement choices. These special cases would be interesting in their own right, and useful for illustrating how the proposal works, but we need a more general strategy to deal with all the rest.

\subsection{Global strategy: the constraint-free regime}
 As we have already noted, ordinary experimental control is time-asymmetric. We control the \textit{initial} conditions of experiments, but not their \textit{final} conditions. Let's call this time-asymmetry \textit{Initial Control} (IC). Some writers call it simply \textit{Causality} \cite{Chiribella10,Coecke14}. 
 
 We want to consider an imaginary regime in which IC is absent. This is easy, so long as we assume, as many who have considered the question have in any case concluded, that the source of this time-asymmetry is the same as that of the thermodynamic asymmetry. In other words, it is the Past Hypothesis.\footnote{For discussion of these issues,  see, for example: \cite{Price96, Albert02,Price10,PriceWeslake10,Rovelli21}.} With this assumption, the regime in which IC is missing is just the big space of possible universes from which the Creator is making a choice in Figure~\ref{fig:god}, when they pick out the one in which PH holds. In other words, it is just what we are left with, when we `turn off' PH itself. 

 So, modulo our assumption that PH is the ultimate source of IC, we have an easy way of considering an imaginary \textit{constraint-free regime} (CFR) in which IC is absent: the assumption guarantees that the PH-free regime is a CFR. And in a CFR, causation and control are time-symmetric by definition, \textit{to the extent that they survive at all.} The qualification is important. If abandoning Initial Control meant abandoning causality altogether, we could no longer speak of colliders, in either temporal direction.

If nothing else, however, the CFR will preserve the two-way determination by boundary conditions that Hawking  has in mind \cite[346]{Hawking94}, when he declares that there is no such thing as a time-asymmetry of causation.\footnote{As philosophers such as \cite{Reich56} have wrongly supposed, in Hawking's view. As we have just seen, Wheeler is equally insistent that causation \textit{is} time-asymmetric. The most likely diagnosis of the disagreement is that Hawking and Wheeler are talking about different things, but we needn't explore this here.}
\begin{quote}
[I]n physics we believe that there are laws that determine the evolution of the universe uniquely. So if state A evolved into state B, one could say that A caused B. But one could equally well look at it in the other direction of time, and say that B caused A. So causality does not define a direction of time. 
\end{quote}
With (a probabilistic version of) this weak notion as a fallback, 
we can say that the CFR permits \textit{some} notion of causality. By considering hypothetical changes to conditions on boundaries, we can also save a weak counterfactual notion of influence, or making a difference. And it will all be time-symmetric, just as Hawking says.

Some readers may be troubled by the presumption that there could be experiments at all in a CFR. But here we are on well-trodden ground: a community familiar with  Boltzmann Brains should not be troubled by a few simple pieces of quantum optics. Such things will arise by chance in a sufficiently expansive CFR, and we may discuss their characteristic statistics.\footnote{We will encounter another example of such thinking, in a QM context, in \S8 below.}

\begin{figure}[t]
\centering
\begin{tikzpicture}
    \node[draw,rectangle,minimum width=0.6cm,minimum height=0.5cm] (I) at (3.1,-2.5) {\textsf{C}};
\node[] (I0) at (2.5,-3.4) {0};
 \node[] (I1) at (2.9,-3.4) {1};
 \node[] (I2) at (3.3,-3.4) {2};
 \node[] (I3) at (3.7,-3.4) {3};
  \node[] (C) at (3.1,-4) {$C$};

  \path [dashed,-] (I1) edge (I);
  \path [dashed,-] (I0) edge (I);
  \path [dashed,-] (I2) edge (I);
  \path [dashed,-] (I3) edge (I);

    \node[draw,rectangle,minimum width=0.6cm,minimum height=0.5cm] (Abox) at (1.2,0) {\textsf{A}};
    \node[] (Alabel) at (1.2,1.7) {$A$};
    \node[] (A0) at (0.8,1) {0};
    \node[] (A1) at (1.6,1) {1};
    \node[] (a) at (0.5,-1) {$a$};
    \node[draw,rectangle,minimum width=0.6cm,minimum height=0.5cm] (Bbox) at (5,0) {\textsf{B}};
      \node[] (Blabel) at (5,1.7) {$B$};
       \node[] (B0) at (4.6,1) {0};
    \node[] (B1) at (5.4,1) {1};
    \node[] (b) at (5.7,-1) {$b$};

   \path [dashed,-] (Abox) edge (A0);
   \path [dashed,-] (Abox) edge (A1);
   \path [->] (a) edge (Abox.south west);
    \path [dashed,-] (Bbox) edge (B0);
   \path [dashed,-] (Bbox) edge (B1);
    \path [->] (b) edge (Bbox.south east);

   \draw [color=blue,line width=1.5pt,-] (I.north) -- (Abox.south east);
   \draw [color=blue,line width=1.5pt,-] (I.north) -- (Bbox.south west);


    \draw[color=olive,line width=1.5pt] (a) to (Abox.south) to (I.north west);
    \draw[color=olive,line width=1.5pt] (b) to (Bbox.south) to (I.north east);

\end{tikzpicture}
\caption{V-shaped case in a CFR} \label{fig:V2}
\end{figure}
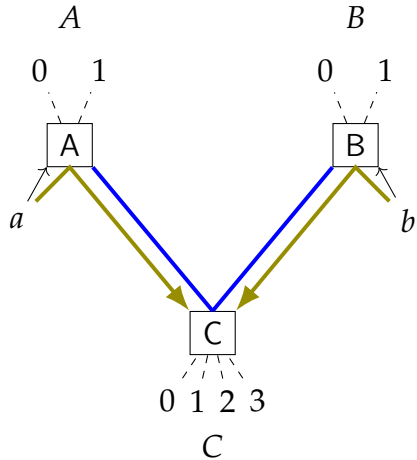

In this framework, then, what does it mean to consider the `effect' of a change to the measurement settings at \textsf{A} and \textsf{B}? It means that we compare the statistics for two classes of random instances of this kind, stipulating as boundary conditions differences with respect to the values of these settings. Because the regime is time-symmetric by definition,  we have no reason to exclude the possibility that the settings `choices' -- the term is highly artificial in this context, of course --  make a difference to the initial state $C$ at \textsf{C}, as shown in olive green in Figure~\ref{fig:V2}. We have changed the colour of the arrows to remind readers that this involves a different sense of causation than in Figure~\ref{fig:V2a}. The initial state thus becomes a collider, and Bell correlations can now be explained as collider artefacts, just as in the delayed-choice W-shaped case.

Specifically, there is no need for a change in Alice's measurement setting $a$ ever to make a difference to Bob's outcome $B$. The difference can always be absorbed as a change in the input state  $C$. Again, this is exactly as in the delayed-choice W case, in its normal version, without a boundary constraint at the collider. The variable at \textsf{C} acts as a safety valve in this case, just like the variable at \textsf{M} in Figure~\ref{fig:W}.

 \subsection{Reimposing Initial Control}

 Whether we take the local or the global path to an unconstrained version of the V-shaped experiment, the normal case in Figure~\ref{fig:V} is exactly what we'd expect, if we reimpose Initial Control at \textsf{C} -- in other words, if we treat the input $C$ as fixed. This closes the safety valve at \textsf{C}, whether the framework in question is that of Figure~\ref{fig:V2a} or Figure~\ref{fig:V2}. 
 
 With the safety valve closed, the Bell correlations now become counterfactually robust. Differences to the setting $a$ must make a difference to the outcome $B$, in some cases (and similarly, of course, from $b$ to $A$).  So the proposed explanation of Bell nonlocality in Figure~\ref{fig:V} works just as well here as in the delayed-choice W case with the Horowitz-Maldacena boundary constraint (Figure~\ref{fig:W2}). Here, too, nonlocality can be explained as CCC, resulting from a combination of collider bias and collider constraint.
 
 We suspect that the reason this parallel hasn't been noticed is that what's normal in the two cases is completely reversed. For W, lack of a boundary constraint is normal, and the constrained case is exceptional. For V, lack of boundary constraint is exceptional, and the constrained case is normal. The logic is exactly the same in both experiments, but to see the parallels we need to put the two exceptional cases on the table, as well as the two normal cases. The result is Table~\ref{fig:VW}, where the greyed-out options are those that are `hard to see', not being manifest in ordinary circumstances.

\begin{table}[ht]
\begin{center}
\resizebox{\textwidth}{!}{%
\begin{tabular}{@{}rll@{}}
\toprule
\multicolumn{1}{l}{}                   & Unconstrained colliders               & Constrained colliders                                     \\ \midrule
\multicolumn{1}{r}{W} & \multicolumn{1}{l}{Normal} & \multicolumn{1}{l}{\textcolor{gray}{Exceptional (only  in black holes?)}} \\ \cmidrule(l){2-3} 
V                      & \textcolor{gray}{Exceptional (only in CFR?)}                 & Normal                                          \\ \bottomrule
\end{tabular}%
}
\end{center}
\caption{Comparison of V and W cases} \label{fig:VW}
\end{table}

The notion of a constrained collider has a remarkable status, if our proposal is correct. It is a very simple idea, apparently playing a ubiquitous role in `the workings of our actual universe', in Penrose's phrase. Yet its sheer ubiquity in one temporal direction, and rarity in the other, has rendered it almost invisible. 


\section{Penrose on quantum measurement}

At this point, we want to compare our proposal to an old claim by Penrose, and to objections to that claim by others. As we'll see, Penrose himself notes that few people find his argument convincing. In our view, the steps needed to see the flaws in his argument are similar to those we have proposed to exhibit the role of IC in V-shaped Bell experiments. For readers familiar with Penrose's argument, then, it may provide a helpful analogy to the present case. 

\begin{figure}[t]
\centering
\includegraphics[width=13.5cm]{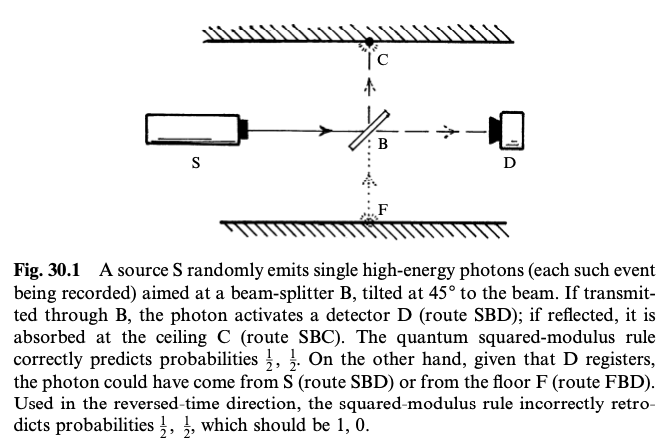}
\caption{\cite[820]{Penrose04} on the apparent time-asymmetry of quantum measurement} \label{fig:pen}
\end{figure}

Penrose has long argued that there is a fundamental time-asymmetry in QM, in the component of the theory relevant to measurement. In \cite{Penrose89,Penrose04}, he discusses the example in Figure~\ref{fig:pen}. He points out that QM predicts a probability $1/2$ that a photon emitted from the source on the left will be registered at the detector on the right; and also probability 1/2 that it will be absorbed in the ceiling, after reflection at the half-silvered mirror. In reverse, however, the probability that a photon detected at the right originated on the left is not $1/2$, but $1$. The probability that it originated instead from the floor below the mirror is effectively $0$.

A common response to Penrose's argument is described by John Baez \cite{Baez96}, who recounts presenting it to Penrose in person. 
\begin{quotation}
\noindent I told him I'd seen this argument and found it very annoying. He said that everyone said so, but nobody had refuted it. \ldots\ I said that \ldots\ if the whole system were in a box and had reached thermodynamic equilibrium, so the walls were just as hot as the lightbulb, we would be justified in concluding, upon seeing a photon at [D], that there was a 50-50 chance of it originating from [S] or [F]. It is only that our world is in a condition of generally increasing entropy that allows for the setup with a hot lightbulb and cool walls to occur, and we can't blame quantum mechanics for that time asymmetry.

[Penrose] thought a while and said that well, a condition of thermal disequilibrium was necessary for a measurement to be made at all, but the real mystery was why we feel confident in using quantum mechanics to predict and not retrodict. This mystery can be traced back to gravity, in that gravity is the root of the arrow of time. \ldots\ I was a bit disappointed that he didn't think my remarks dealt a crushing blow to this thought experiment, but I had to agree that if this was all he thought the moral of the experiment was, he was right.
\end{quotation}

Baez was right to be disappointed, in our view. The need to take into account the low entropy past in retrodiction is not some peculiar feature of QM. On the contrary, it is universal, as our discussion of the Past Hypothesis in \S6 noted. Sean Carroll puts the point like this, for example:   
\begin{quote}
    Ordinarily, when we have some statistical system, we know some macroscopic facts about it but only have a probability distribution over the microscopic details. If our goal is to predict the future, it suffices to choose a distribution that is uniform in the Liouville measure given to us by classical mechanics (or its quantum analogue). If we want to reconstruct the past, in contrast, we need to conditionalize over trajectories that also started in a low-entropy past state. \cite{Carroll13}
\end{quote}

Interpreted as Penrose describes it to Baez, then, the point has nothing specifically to do with QM. Yet some years after that conversation, in \cite{Penrose04}, Penrose continues to present the point as a distinctive observation about QM. Referring to objections such as Baez's, he says this:
\begin{quote}
Sometimes people have objected to this deduction, pointing out that I have failed to take into account all sorts of particular circumstances that pertain to my time-reversed description, such as the fact that Second Law of thermodynamics only works one way in time, or the fact that the temperature of the floor is much lower than that of the source, etc. But the wonderful feature of the quantum-mechanical squared-modulus law is that we never have to worry about what the particular circumstances might be! The miracle is that the quantum probabilities for future predictions arising in the measurement process do not seem to depend at all on considerations of particular temperatures or geometries or anything. If we know the amplitudes, then we can work out the future probabilities. All we need to know are the amplitudes. The situation is completely different for the probabilities for retrodiction. Then we do need to know all sorts of detailed things about the circumstances. The amplitudes alone are quite insufficient for computing past probabilities. \cite[821]{Penrose04}
\end{quote}
Again, however, the same is true of classical statistical probabilities, as Carroll's remark above makes clear. 

Penrose adds an endnote to this discussion, that takes us in the direction of our present concerns. 
\begin{quote}
I find it remarkable how much difficulty people often have with this argument. The matter is perhaps clarified if we contemplate numerous occurrences of \textit{this experiment,} taking place at various locations throughout spacetime. There are four alternative photon routes to be considered, SBD, SBC, FBD, and FBC. To see what the various probabilities are, we ask for the proportion of SBD, given S (forward-time situation), or for the proportion of SBD, given D (backward-time situation). The squared-modulus rule correctly gives the actual answer (50\%) in the first case, but it does not give the actual answer (nearly 100\%) in the second case. \cite[866, emphasis added]{Penrose04} 
\end{quote}
However, everything hangs on what we mean by an occurrence of `this experiment'. If we stipulate that it means that the photon is at S at the beginning of the experiment, then we have put the asymmetry that Penrose wants in by hand, and it no surprise that we can extract it again later. But if remove this stipulation for the reverse cases, stipulating only that the photon is at D at the end of the experiment, then we get Baez's conclusion. So long as we also equalise for the thermodynamic factors -- which we might do, in principle, by considering our constraint-free regime, where PH is turned off, and the experiment in question occur as fluctuations in a universe in thermodynamic equilibrium overall -- then there is simply no asymmetry in the frequencies of the kind that Penrose requires.\footnote{The imaginative exercise here is similar to that required in our global route to an unconstrained version of our V-shaped Bell experiment, as in \S7.2 above.}

This example has several lessons for our discussion. \begin{enumerate}
    \item It illustrates how much trouble we can have in setting aside Initial Control. Penrose's basic mistake is to ignore the role it is playing, even in his presentation in the last passage quoted above, supposedly generalised to `various locations throughout spacetime'.
    \item It demonstrates that we can clarify matters by considering an imaginary regime in which Initial Control is absent -- but that we need to do it properly, and not simply imagine a lot of cases in which it is present!
    \item It shows that it is worth making the effort to consider this imaginary regime, though for different reasons in the two cases. In Penrose's case, it is the key to avoiding a fallacious argument about a fundamental time-asymmetry in quantum measurement. In the present case, it allows us to see that entanglement may be a product of constrained colliders. IC provides the constraint, but in order to see the role it is playing,  we need to imagine it absent.
    \item Finally, the analogy also illustrates that there may be less drastic paths to the relevant conclusion, in both cases. It seems easy to construct an analogue of our Milburn experiment in \S7.1, for Penrose's case: we just need a second source on the floor at F, apparently. This is comparable to our local route to an unconstrained version of the V-shaped experiment. So in both cases, we may not need the full resources of the CFR, to see the role that IC is playing in ordinary situations.
\end{enumerate}

 \section{Does the proposal generalize?}

We have proposed a mechanism  to account for Bell correlations in certain important cases, but how far does it generalise? The fact that it deals with some of the canonical entanglement experiments is a reason for optimism, but can we do better? Alternatively, can we show that it does \textit{not} generalize to some entangled systems? 

The important question seems to be this. Given an arbitrary entangled system S, can we always find a natural `super-ensemble' \textbf{S} in our CFR, containing S as one of its members, with the following properties?

\begin{enumerate}[label=(\roman*)]
    \item In the natural time-symmetric measure on the CFR, \textbf{S} as a whole `washes out' the correlations taken to be manifestations of  entanglement in S in the first place. \item We can restore these original correlations by imposing the actual initial state of S as a boundary condition. \end{enumerate} 

For cases in which the entangled state in question has an initial preparation, as in the V-shaped examples, this seems to be straightforward. In QM it is standard to use a `density matrix', $\rho$, to represent states with an uncertain preparation.  In this framework, it is easy to show that a completely uncertain preparation (over a sufficiently large number of possibilities) will always result in what is known as a `maximally mixed state'.  Such states have exactly the properties required here; there are no correlations which can be observed in a maximally-mixed state.  Yet, if one selects one particular preparation state $|\psi\rangle$, the density matrix must take the form $\rho=|\psi\rangle\langle\psi |$, which can never be maximally mixed.  So quantum correlations can always be restored by exercising initial control on a prepared quantum system. 

Conversely, removing this initial control -- replacing it with an equal-probability set of preparations which span the Hilbert space -- always removes these correlations.  This makes the initial control entirely responsible for the eventual correlations, for any jointly-prepared entangled state. 

However, there are further ways to generate entangled states, different from both the V- and W-shaped examples. Examples include the Hong-Ou-Mandel effect \cite{Hong87}, and so-called `which-way' entanglement \cite{Ferrari10}.\footnote{Though the latter cases, at least, can always be reduced to more conventional entanglement scenario, where each path can be recast as an entangled state of zero-particle and one-particle modes \cite{catani23}.}   As it stands, the present proposal does not capture these cases.  

 \section{Summary}

We have argued that in both V-shaped and W-shaped cases, Bell nonlocality can result from a combination of two things: (i) causal influences operating within the lightcones; and (ii) boundary constraints, restricting the possible values of collider variables. This turns out to be capable of  producing a counterfactual-supporting spacelike connections, even though neither of the individual ingredients is in tension with special relativity. It has long been recognised that in principle, zigzag influences might render EPR phenomena compatible with relativity.\footnote{See, for example, \cite{Costa53,Price96,PriceWharton15,WhartonArgaman20}.} This proposal offers a mechanism to produce such zigzag connections. 

Our argument rests on a bridge between  two familiar sets of ideas. On one side, it is standard fare of causal modelling that {conditioning on colliders} produces selection artefacts. If we apply these ideas in QM, it turns out to be uncontroversial that \textit{some} Bell correlations (i.e., those in DCES W-shaped experiments) are properly explained this way.\footnote{As in \S4.1, it is easy to show that postselection can produce Bell correlations in toy models. The interesting fact is that it actually happens at a fundamental level in QM.\label{fn:toy}} The notion of \textit{constraining} a collider is not standard fare, but once we have it, we can find at least one version of it in physics, in the Horowitz-Maldacena proposal.

 On the other side, it is a familiar idea that most (at least) of  the  ordinary time-asymmetry we observe in our universe results from a low-entropy boundary condition on a time-symmetric space of possible histories -- from \textit{conditioning on the Past Hypothesis} (PH), as  \cite{Albert02} puts it. We noted that `conditioning' is here comparable not to ordinary conditioning on colliders, where real instances need to be discarded, but to the constrained case, where the constraint ensures that such instances don't happen in the first place. In order to condition on PH, we do not have to ignore a vast number of \textit{actual} universes in which PH does not hold.  

So `preconstraint' is already at work in our universe, on a massive scale. To connect its existence to constrained colliders, and to the origin of Bell correlations in ordinary V-shaped cases, we needed to imagine unconstrained cases of those experiments. For this we proposed two routes, both of which bring into view the possibility of unconstrained V-shaped cases. Once such cases are in view, it is easy to see how ordinary V-shaped Bell correlations can \textit{{also}} be understood as CCC, with constraint achieved in the familiar ways permitted by Initial Control. Within the correlational structures dictated by QM, Bell nonlocality itself\footnote{In the indirect, zigzag form that CCC permits.} seems to be another of the local, downstream manifestations of the Past Hypothesis.

\subsection{Open questions}
In \S1  we mentioned Penrose's `two mysteries of quantum entanglement'. If successful, our proposal does something to address the first mystery: `How \ldots\ to come to terms with quantum entanglement and to make sense of it in terms of ideas that we can comprehend', as Penrose puts it  \cite[591]{Penrose04}   
This turns out to be surprisingly simple, if the proposal above is correct. The ingredients needed are simply collider bias, collider constraint, and Initial Control. The first and third ingredients are familiar, while the second is also available off the shelf, once we know how to ask for it. All three are readily comprehensible, and -- given the  the role of the Past Hypothesis -- ubiquitous in our actual universe. 

However, this simplicity and ubiquity seems to make Penrose's second question about entanglment even more pressing: 
\begin{quote}
[W]hy is it something that we barely notice in our direct experience of the world? Why do these ubiquitous  effects of entanglement not confront us at every turn?\end{quote} 
Penrose adds that he does `not believe that this second mystery has received nearly the attention that it deserves, people’s puzzlement having been almost entirely concentrated on the first'  \cite[591]{Penrose04}.

 To close, we note a third question, related but not identical to Penrose's second. If this proposal is correct, then why does the simple mechanism it describes seem to be confined to the quantum world?\footnote{At least at a fundamental level; cf.~fn.~\ref{fn:toy}.} This question has two parts to it. Why does is this mechanism active in the quantum world at all; and why, apparently, nowhere else? To readers who find the main proposal plausible, we suggest these questions, and the issue of generality from \S9, as interesting topics for further work.\footnote{We are grateful to Gerard Milburn, Michael Cuffaro, Jeff Bub, and Heinrich P\"as,  
 for discussion and helpful comments on related material.}


\begin{thebibliography}{999}



\bibitem[Albert 2000]{Albert02}Albert, D. \textit{Time and Chance,} Cambridge, MA: HUP.



\bibitem[Bacciagaluppi \& Hermens 2021]{Guido21} Bacciagaluppi, G. \& Hermens, R. Bell inequality violation and relativity of pre- and postselection. arXiv:2002.03935

\bibitem[Baez 1996]{Baez96} Baez, John.  A chat with Penrose, June 10, 1996. Blog post. Accessed 17 May 2024 at \href{https://math.ucr.edu/home/baez/penrose.html}{math.ucr.edu/home/baez/penrose.html}.



\bibitem[Bell 1964]{Bell64} Bell, J. S. On the Einstein-Podolsky-Rosen paradox. \textit{Physics,} 1, 195--200, reprinted in \cite{Bell04}.

\bibitem[Bell 2004]{Bell04} Bell, J.~S. \textit{Speakable and Unspeakable in Quantum Mechanics, Second Edition.} Cambridge University Press: Cambridge.

\bibitem[Berkson 1946]{Berkson46} Berkson, J. Limitations of the application of fourfold table analysis to hospital data. \textit{Biometrics Bulletin,}  2,  47--53.


\bibitem[Bohm 1951]{Bohm51}Bohm, D. \textit{Quantum Theory}. Englewood Cliffs, NJ: Prentice-Hall.



\bibitem[Carroll 2013]{Carroll13}Carroll, Sean M. 2013. Cosmology and the Past Hypothesis. Blog post. Accessed 10 May 2024 at: \href{https://www.preposterousuniverse.com/blog/2013/07/09/cosmology-and-the-past-hypothesis/comment-page-2/}{www.preposterous\-universe.com/blog/2013/07/09/cosmology-and-the-past-hypo\-the\-sis/comment-page-2/} 

\bibitem[Catani et al 2023]{catani23}Catani, L., Leifer, M., Schmid, D.~and Spekkens, R. W. (2023). Why interference phenomena do not capture the essence of quantum theory. \textit{Quantum} 7, 1119. arXiv:2111.13727 [quant-ph]

\bibitem[Chiribella et al 2010]{Chiribella10} G. Chiribella, G. M. D’Ariano, and P. Perinotti. Probabilistic theories with purification. \textit{Physical Review A} 81(6), 062348.

\bibitem[Coecke 2014]{Coecke14} Coecke, Bob. 2014. Terminality implies non-signalling. arXiv:1405.3681.


\bibitem[Cole et al 2010]{Cole10} Cole, S., Platt, R., Schisterman, E., Chu, H., Westreich, D., Richardson, D., Poole, C. Illustrating bias due to conditioning on a collider. \textit{International Journal of Epidemiology,} 39,  417--420. https://doi.org/10.1093/ije/dyp334

\bibitem[Costa de Beauregard 1953]{Costa53} Costa de Beauregard, O. M\'echanique quant\-ique. \emph{Comptes Rendus Acad\'emie des Sciences} 236, 1632--34.

\bibitem[Czeisler et al 2021]{Czeisler21}Czeisler, M.É., Wiley, J.F., Czeisler, C.A., Rajaratnam, S.M.W.~and Howard, M.E. Uncovering survivorship bias in longitudinal mental health surveys during the COVID-19 pandemic. \textit{Epidemiol Psychiatr Sci.,} 30:e45. doi: 10.1017/S204579602100038X.

 \bibitem[Egg 2013]{Egg13} Egg, M. Delayed-choice experiments and the metaphysics of entanglement. \textit{Foundations of Physics,} 43, 1124--1135.

\bibitem[EPR 1935]{EPR}
Einstein, A., Podolsky, B.~and Rosen, N. Can quantum-mechanical description of physical reality be considered complete? \textit{Physical Review,} 47, 777–780.




\bibitem[Fankhauser 2019]{Fankhauser19} Fankhauser, J. Taming the delayed choice quantum eraser. \textit{Quanta,} 8, 44--56. arXiv:1707.07884


 	


\bibitem[Ferrari \& Braunecker 2010]{Ferrari10}Christian Ferrari, Bernd Braunecker. Entanglement, which-way measurements, and a quantum erasure. \textit{Am. J. Phys.} 78: 792–795. https://doi.org/10.1119/1.3369921

\bibitem[Friedrich \& Evans 2019]{FriedrichEvans19} Friederich, S. and Evans, P. Retrocausality in quantum mechanics. In \textit{The Stanford Encyclopedia of Philosophy} (Summer 2019 Edition); Zalta, Edward (ed.). \href{http://plato.stanford.edu/archives/sum2019/entries/qm-retrocausality/}{http://plato.stanford.edu/archives/sum2019/entries/qm-retro\-causality/}

\bibitem[Frisch 2007]{Frisch07}Frisch, Mathias. Does a Low-Entropy Constraint Prevent Us from Influencing the Past? In G.~Ernst 
and Andreas H\"uttemann(eds), \textit{Time, chance and reduction: philosophical aspects of statistical mechanics.} New York:Cambridge University Press. pp.~13--33.



\bibitem[Gaasbeek 2010]{Gaasbeek10} Gaasbeek, B. Demystifying the delayed choice experiments. arXiv:1007.3977

\bibitem[Gibbard \& Harper 1978]{Gibbard78}
Gibbard, A.~and Harper, W. Counterfactuals and two kinds of expected utility. In C.
Hooker, J. Leach and E. McClennen (eds.), \textit{Foundations and Applications of Decision Theory,}
Dordrecht: Reidel, 125--162.

\bibitem[Gömöri \& Hoefer 2023]{Gomori23}
Gömöri, M. \& Hoefer, C. Classicality and Bell’s theorem. \textit{European Journal for Philosophy of Science,} 13, 45. https://doi.org/10.1007/s13194-023-00531-y


\bibitem[Giustina et al 2015]{Giustina15} Giustina, M., Versteegh, M.A., Wengerowsky, S., Handsteiner, J., Hochrainer, A., Phelan, K., et al. Significant-loophole-free test of Bell’s theorem with entangled photons. \textit{Physical review letters,} 115(25), 250401. arxiv:1511.03190


\bibitem[Hawking 1994]{Hawking94} Hawking, S. The no boundary condition and the arrow of time.  In Halliwell, Perez-Mercader, and Zurek (eds), \textit{Physical Origins of Time
Asymmetry,} Cambridge University Press, 346--357.


\bibitem[Hensen et al 2015]{Hensen15} Hensen, B., Bernien, H., Dreau, A. E., Reiserer, A., Kalb, N., Blok, M. S., et al. Loophole-free Bell inequality violation using electron spins separated by 1.3 kilometres. \textit{Nature,} 526, 682--686. arXiv:1508.05949


\bibitem[Hofer-Szabó et al 2013]{Hofer2013} Hofer-Szabó, G., Rédei, M., \& Szabó, L. \textit{The Principle of the Common Cause.} Cambridge: Cambridge University Press. doi:10.1017/CBO9781139094344

\bibitem[Hong et al 1987]{Hong87} Hong, C.K., Ou, Z.Y.~and Mandel, L. Measurement of subpicosecond time intervals between two photons by interference. \textit{Phys. Rev. Lett.} 59, 2044.

\bibitem[Horowitz \& Maldacena 2004]{HorowitzMaldacena04}
Horowitz, G.~and Maldacena, J. The black hole final state, \textit{JHEP} 0402:008. arXiv:hep-th/0310281

\bibitem[Hossenfelder \& Palmer 2020]{Hoss19} Hossenfelder, S. \& Palmer, T. Rethinking superdeterminism. \textit{Frontiers of Physics,} 06 May 2020. doi.org/10.3389/fphy.2020.00139. arXiv:1912.06462 








\bibitem[Ma et al 2012]{Ma12} Ma, X.-s., Zotter, S., Kofler, J., Ursin, R., Jennewein, T., Brukner, C., et al. Experimental delayed-choice entanglement swapping. \textit{Nature Physics,} 8, 479--484. arXiv:1203.4834

\bibitem[Maudlin 2011]{Maud11} Maudlin, T. \textit{Quantum Non-Locality and Relativity: Metaphysical Intimations of Modern Physics} (3rd edn.). Oxford: Basil Blackwell.

\bibitem[Maudlin 2014]{Maud14}Maudlin, T. What Bell did. \textit{J.~Phys.~A: Math.~Theor.,} 47, 424010.
DOI 10.1088/1751-8113/47/42/424010

\bibitem[Mjelva 2024]{Mjelva24}Mjelva, J{\o}rn. Delayed-choice entanglement swapping experiments: no evidence for timelike entanglement. Forthcoming in \textit{Studies in History and Philosophy of Science.} \href{https://philpapers.org/rec/MJEDES}{philpapers.org\-/rec/MJEDES}


\bibitem[Myrvold et al 2021]{Myrvold21} Myrvold, W., Marco G. \& Shimony, A. Bell’s Theorem. \textit{The Stanford Encyclopedia of Philosophy (Fall 2021 Edition),} Zalta, E. (ed.). \href{https://plato.stanford.edu/archives/fall2021/entries/bell-theorem/}{https://plato.stanford.edu/archives/fall2021/entries/bell-theorem/}




\bibitem[Penrose 1989]{Penrose89}Penrose, Roger. \textit{The Emperor’s New Mind: Concerning Computers, Minds, and The Laws of Physics.} Oxford University Press.

\bibitem[Penrose 2004]{Penrose04}Penrose, Roger. \textit{The Road to Reality.} London: Jonathan Cape.



\bibitem[Perry 2021a]{Perry21}
Perry, M. No future in black holes. arXiv:2106.03715


\bibitem[Perry 2021b]{Perry21b} Perry, M. Future Boundaries and the Black Hole Information Paradox. arXiv:2108.05744.


\bibitem[Pigou 1911]{Pigou11} Pigou, A.~C. Alcoholism and heredity. \textit{Westminster Gazette,} 2nd February 1911. Reprinted in \textit{International Journal of Epidemiology,} 51(2022), e227--e228. doi.org/10.1093/ije/dyw340


\bibitem[Price 1996]{Price96} Price, Huw. \textit{Time's Arrow and Archimedes' Point,} Oxford University Press, New York.


\bibitem[Price 2010]{Price10}Price, Huw. Time's arrow and Eddington's challenge. \textit{Séminaire Poincaré XV, Le Temps,} 115--140. Accessible  at \href{http://www.bourbaphy.fr/price.pdf}{www.bourbaphy.fr/\-price.pdf}

\bibitem[Price 2012]{Price12} Price, Huw. Does time-symmetry imply retrocausality? How the quantum world says ``maybe''. \textit{Studies in History and Philosophy of Modern Physics,} 43, 75--83. arXiv:1002.0906 

\bibitem[Price 2024]{Price24} Price, Huw. W as the edge of a wedge: Bell correlations via constrained colliders. arXiv:2404.13928 [quant-ph]

\bibitem[Price \& Weslake 2010]{PriceWeslake10}
Price, Huw \& Weslake, Brad. The time-asymmetry of causation. In Helen Beebee, Christopher Hitchcock and Peter Menzies (eds), \textit{The Oxford Handbook of Causation} (OUP), 414--443.

\bibitem[Price \& Wharton 2015]{PriceWharton15} Price, Huw \& Wharton, Ken. Disentangling the quantum world. \textit{Entropy} 17:11, 7752--7767. arXiv:1508.01140

\bibitem[Price \& Wharton 2016]{PriceWharton16} Price, Huw \& Wharton, Ken. Taming the quantum spooks. \textit{Aeon}, 14 September 2016. \href{https://aeon.co/essays/can-retrocausality-solve-the-puzzle-of-action-at-a-distance}{https://aeon.co/essays/can-retrocausality-solve-the-puzzle-of-action-at-a-distance}


\bibitem[Price \& Wharton 2021a]{PriceWharton21a}
Price, Huw \& Wharton, Ken. Entanglement swapping and action at a distance. \textit{Foundations of Physics,} 51, 105. doi.org/10.1007/s10701-021-00511-3 

\bibitem[Price \& Wharton 2021b]{PriceWharton21b}
Price, Huw \& Wharton, Ken. Appendix to ArXiV version of \cite{PriceWharton21a}.  arXiv:2101.05370v4 [quant-ph]



\bibitem[Price \& Wharton 2022]{PriceWharton22}
Price, Huw \& Wharton, Ken. Why entanglement?  arXiv:2212.06986 [quant-ph]. 
doi.org/10.48550/arXiv.2212.06986


\bibitem[Price \& Wharton 2023]{PriceWharton23}
Price, Huw \& Wharton, Ken. Untangling entanglement. \textit{Aeon}, 29 June 2023. \href{https://aeon.co/essays/our-simple-magic-free-recipe-for-quantum-entanglement}{https://aeon.co/essays/our-simple-magic-free-recipe-for-quantum-entanglement}

\bibitem[Reichenbach 1956]{Reich56} Reichenbach, Hans. \textit{The direction of time.} Edited by Maria Reichenbach. Mineola, N.Y.: Dover Publications. 


\bibitem[Rovelli 2021]{Rovelli21} 
Rovelli, C. Agency in physics. In Claudio Calosi, Pierluigi Graziani, Davide Pietrini, Gino Tarozzi, eds.,  \textit{Experience, abstraction and the scientific image of the world: Festschrift for Vincenzo Fano.} Franco Angeli editore. Available at arXiv:2007.05300.


\bibitem[Shalm et al 2015]{Shalm15} Shalm, L.K., Meyer-Scott, E., Christensen, B.G., Bierhorst, P., Wayne, M.A., Stevens, M.J., et al. Strong loophole-free test of local realism. \textit{Physical Review Letters,} 115(25), 250402. arxiv:1511.03189


\bibitem[Schrödinger 1935a]{Sch35a} Schrödinger, E. Die gegenwärtige Situation in der Quantenmechanik. \textit{Naturwissenschaften,} 23, 807--812; English translation in Trimmer, 1980. doi:10.1007/BF01491891

\bibitem[Schrödinger 1935b]{Sch35b} Schrödinger, E. Discussion of probability relations between separated systems. \textit{Mathematical Proceedings of the Cambridge Philosophical Society,} 31, 555--563.

\bibitem[Trimmer 1980]{Trim80}
Trimmer, J. D. The present situation in quantum mechanics: A translation of Schrödinger’s ‘cat paradox’ paper. \textit{Proceedings of the American Philosophical Society,} 124: 3230--338.



\bibitem[van Fraassen 1982]{vanF82}
 van Fraassen, Bas C. The charybdis of realism: epistemological implications of Bell’s inequality. \textit{Synthese,} 52(1): 25--38. doi:10.1007/BF00485253




\bibitem[Wharton \& Argaman 2020]{WhartonArgaman20}
Wharton, K. \& Argaman, N. Bell's Theorem and locally-mediated reformulations of quantum mechanics. \emph{Reviews of Modern Physics,} \textit{92,} 21002. arXiv:1906.04313

\bibitem[Wheeler 1978]{Wheeler78}Wheeler, J. The `Past' and the `Delayed Choice' Double-Slit Experiment.
In Marlow, A.~(ed.), \textit{Mathematical Foundations of Quantum Theory,} New York:
Academic Press, 9--48.

\bibitem[Wiseman \& Cavalcanti 2017]{Wise17}
Wiseman, H. \& Cavalcanti, E. \textit{Causarum investigatio} and the two Bell's theorems of John Bell. In Bertlmann and Zeilinger (eds.), \textit{Quantum [Un]Speakables II: Half a Century of Bell's Theorem} (Springer, Switzerland), 119--142. arXiv:1503.06413 [quant-ph]


\bibitem[Wood \& Spekkens 2015]{Wood15} Wood, C.~J.~\& Spekkens, R.~W.
The lesson of causal discovery algorithms for quantum correlations: causal explanations of Bell-inequality violations require fine-tuning. 
\textit{New Journal of Physics,} 17. arXiv:1208.4119 [quant-ph]. DOI 10.1088/1367-2630/17/3/033002




\end{thebibliography}
\end{document}